\documentclass[a4paper,oneside,final,notitlepage,onecolumn,12pt]{article}
\usepackage{amsfonts}
\usepackage{epsf}
\usepackage{graphicx}
\usepackage{amssymb,eso-pic}
\usepackage{latexsym}
\usepackage{tabularx}
\usepackage{amsxtra} 
\usepackage{hyperref}
\usepackage{t1enc}
\usepackage{amsmath,accents}
\usepackage{bbm}
\usepackage{enumerate}
\usepackage{cancel}

\usepackage{ifpdf,epsfig,array,amsmath,amssymb}

\usepackage{psfrag} 
\psfrag{na}{\footnotesize $n^a$}   
\psfrag{ta}{\footnotesize $\sigma^a$}   
\psfrag{vna}{\footnotesize $\check{n}^a$}
\psfrag{vNa}{\footnotesize $\check{N}^a$}
\psfrag{scrW}{\footnotesize $\mycal{W}_{\rho_0}$}
\psfrag{t=t1}{\footnotesize $\Sigma_{\sigma_1}$}
\psfrag{t=t2}{\footnotesize $\Sigma_{\sigma_2}$}
\psfrag{r=all}{\footnotesize $\rho=const$}
\psfrag{hna}{\footnotesize $\,\widehat{n}^a$}
\psfrag{tna}{\footnotesize $\tilde{n}^a$}
\psfrag{hab,Kab}{\footnotesize $h_{ab}, K_{ab}$}
\psfrag{thab,tKab}{\footnotesize $\tilde h_{ab}, \tilde K_{ab}$}
\psfrag{hhab,hKab}{\footnotesize $\widehat h_{ab}, \widehat K_{ab}$}
\psfrag{chab,cKab}{\footnotesize $\check{h}_{ab}, \check{K}_{ab}$}

\usepackage[normalem]{ulem}
\usepackage{color}
\definecolor{blue}{rgb}{0,0,1}
\definecolor{red}{rgb}{1,0,0}

\definecolor{DGREEN}{rgb}{0,0.7,0.3}
\definecolor{grey1}{rgb}{0.52, 0.52, 0.51}

\newcommand{\interior}[1]{\accentset{\smash{\raisebox{-0.1ex}{$\scriptstyle\circ$}}}{#1}\rule{0pt}{2.3ex}}
\newcommand{\inbullet}[1]{\accentset{\smash{\raisebox{-0.1ex}{$\scriptstyle\bullet$}}}{#1}\rule{0pt}{2.3ex}}
\newcommand{\indiamond}[1]{\accentset{\smash{\raisebox{-0.1ex}{\rotatebox[origin=c]{90}{$\scriptscriptstyle\lozenge$}}}}{#1}\rule{0pt}{2.3ex}}
\newcommand{\inblacklozenge}[1]{\accentset{\smash{\raisebox{-0.1ex}{\rotatebox[origin=c]{90}{$\scriptscriptstyle\blacklozenge$}}}}{#1}\rule{0pt}{2.3ex}}


\makeatletter
\def\@xfootnote[#1]{%
  \protected@xdef\@thefnmark{#1}%
  \@footnotemark\@footnotetext}
\makeatother

\DeclareFontFamily{OT1}{rsfs}{} \DeclareFontShape{OT1}{rsfs}{m}{n}{
<-7> rsfs5 <7-10> rsfs7 <10-> rsfs10}{}
\DeclareMathAlphabet{\mycal}{OT1}{rsfs}{m}{n}

\def\sc{{\hskip 3.5pt {{}^{{}^{{}_{{}_{\bowtie}}}}} \kern -8.pt{}}}  
\def\SC{{\hskip 3.5pt {{}^{{}^{{}^{{}_{{}_{\bowtie}}}}}} \kern -10.5pt{}}}

\def\d{{\rm d}}

\DeclareMathAlphabet{\mathpzc}{OT1}{pzc}{m}{it}

\newcommand{\hoch}[1]{$\, ^{#1}$}

\newcommand{\auth}{  Istv\'an R\'{a}cz\hoch{1}
and \ Jeffrey Winicour\hoch{2}
}

\begin{document}

\newtheorem{theorem}{Theorem}[section]
\newtheorem{lemma}{Lemma}[section]
\newtheorem{proposition}{Proposition}[section]
\newtheorem{corollary}{Corollary}[section]
\newtheorem{conjecture}{Conjecture}[section]
\newtheorem{example}{Example}[section]
\newtheorem{definition}{Definition}[section]
\newtheorem{remark}{Remark}[section]
\newtheorem{exercise}{Exercise}[section]
\newtheorem{axiom}{Axiom}[section]
\renewcommand{\theequation}{\thesection.\arabic{equation}}

\begin{center}


{\LARGE{\bf On solving the constraints by integrating a strongly hyperbolic system} }

\vspace{25pt}
\auth

\vspace{10pt}{\hoch{1}\it Wigner Research Center for Physics} \\  {H-1121 Budapest, Hungary}

\vspace{10pt}{\hoch{2}\it Department of Physics and Astronomy,\\ 
University of Pittsburg, Pittsburgh, PA, 15260,  USA}

\vspace{15pt}

\today

\begin{abstract}

It was shown recently that the constraints on the initial data for Einstein's equations may be posed as an evolutionary problem \cite{racz_constraints}.
In one of the proposed two methods the constraints can be replaced by a first order symmetrizable hyperbolic system and a
subsidiary algebraic relation. Here, by assuming that the initial data surface is smoothly foliated by a one-parameter family of topological two-spheres,
the basic variables are recast in terms of spin-weighted fields. This allows one to replace all the angular derivatives in the evolutionary system
by the Newman-Penrose $\eth$ and $\overline\eth$ operators which, in turn, opens up a new avenue to solve the constraints by integrating the
resulting system using suitable numerical schemes. In particular, by replacing the $\eth$ and $\overline\eth$ operators either by a finite difference or by a pseudo-spectral representation or by applying a spectral
decomposition in terms of spin-weighted spherical harmonics,
the evolutionary equations may be put into the form of a coupled system of non-linear ordinary differential equations.

\end{abstract} 

\end{center}

\section{Introduction}\label{introduction}
\setcounter{equation}{0}

This paper is intended to be a technical report providing a firm analytic background to support the numerical integration of the Einstein constraint equations,
when cast into the form of a first order symmetrizable hyperbolic system and a subsidiary algebraic condition. The main steps in deriving
this form of the constraint equations is outlined in this section.

\medskip

Consider first the initial data specification in general relativity comprised by a Riemannian metric $h_{ij}$ and a symmetric tensor field $K_{ij}$
on a three-dimensional manifold $\Sigma$. The pair $(h_{ij},K_{ij})$ is said to satisfy the vacuum constraints
(see e.g.~Refs.~\cite{choquet,wald}) if the relations  
\begin{align} 
{}^{{}^{(3)}}\hskip-1mm R + \left({K^{j}}_{j}\right)^2 - K_{ij} K^{ij} =0\,, \label{new_expl_eh}\\
D_j {K^{j}}_{i} - D_i {K^{j}}_{j} =0\label{new_expl_em}
\end{align}
hold on $\Sigma$, where ${}^{{}^{(3)}}\hskip-1mm R$ and $D_i$ denote the scalar curvature and the
covariant derivative operator associated with $h_{ij}$, respectively. 

\medskip

$\Sigma$ is assumed to be smoothly foliated by a one-parameter family of topological two-spheres $\mycal{S}_\rho$
which may also be considered as the level surfaces $\rho=const$ of a smooth function $\rho: \Sigma \rightarrow \mathbb{R}$.

\medskip

Applying a vector field $\rho^i$ on $\Sigma$, satisfying the relation $\rho^i \partial_i \rho=1$, the unit normal $\widehat n^i$
to the level surfaces $\mycal{S}_\rho$ decomposes as
\begin{equation}\label{nhat}
\widehat n^i={\,\widehat{N}}^{-1}\,[\, \rho^i-{\widehat N}{}^i\,]\,,
\end{equation}
where the `lapse' $ \widehat N$ and `shift' $\widehat N^i$ of the vector field $\rho^i$
are determined by
$\widehat n_i= \widehat N \partial_i \rho$ and $\widehat N^i=\widehat \gamma{}^i{}_j\,\rho^j$,
where $\widehat \gamma{}^i{}_j=\delta{}^i{}_j-\widehat n{}^i\widehat n_j$. 

\medskip

The Riemannian metric $h_{ij}$ on $\Sigma$ can then be decomposed as 
\begin{equation}\label{hij}
h_{ij}=\widehat \gamma_{ij}+\widehat  n_i \widehat n_j\,,
\end{equation}
where $\widehat \gamma_{ij}$ is the metric induced on the surfaces $\mycal{S}_\rho$, while 
the extrinsic curvature $\widehat K_{ij}$ of $\mycal{S}_\rho$ is given by
\begin{equation}\label{hatextcurv}
\widehat K_{ij}= {{\widehat \gamma}^{l}}{}_{i}\, D_l\,\widehat n_
j=\tfrac12\,\mycal{L}_{\widehat n} {\widehat \gamma}_{ij}\,.
\end{equation}

\medskip

The other part of the initial data represented by the symmetric tensor field $K_{ij}$ has decomposition
\begin{equation}
K_{ij}= \boldsymbol\kappa \,\widehat n_i \widehat n_j  + \left[\widehat n_i \,{\rm\bf k}{}_j  
+ \widehat n_j\,{\rm\bf k}{}_i\right]  + {\rm\bf K}_{ij}\,,
\end{equation}
where $\boldsymbol\kappa= \widehat n^k\widehat  n^l\,K_{kl}$,
${\rm\bf k}{}_{i} = {\widehat \gamma}^{k}{}_{i} \,\widehat  n^l\, K_{kl}$
and ${\rm\bf K}_{ij} = {\widehat \gamma}^{k}{}_{i} {\widehat \gamma}^{l}{}_{j}\,K_{kl}$. 
Note that all boldfaced symbols stand for tensor fields which are well-defined on the individual leaves $\mycal{S}_\rho$.
In recasting the Hamiltonian and momentum constraints (\ref{new_expl_eh}) and (\ref{new_expl_em}) the traces 
\begin{equation}
{\widehat K}{}^l{}_{l}=\widehat\gamma^{kl}\,{\widehat K}_{kl} \quad {\rm and} \quad {\rm\bf K}^l{}_{l}=\widehat\gamma^{kl}\,{\rm\bf K}_{kl}
\end{equation}
and the trace free part of ${\rm\bf K}_{ij}$, defined as
\begin{equation}\label{intK}
\interior{\rm\bf K}_{ij}={\rm\bf K}_{ij}-\tfrac12\,\widehat \gamma_{ij}\,{\rm\bf K}^l{}_{l} \, ,
\end{equation}
will also be involved. 

\medskip

By making use of the above variables, the pair $(h_{ij},K_{ij})$ may be replaced by the fields 
$\widehat N,\widehat N^i,\widehat \gamma_{ij}, \interior{\rm\bf K}_{ij}, \boldsymbol\kappa,{\rm\bf k}{}_{i}$ and ${\rm\bf K}^l{}_{l}$,
and in turn, the Hamiltonian and momentum constraints (\ref{new_expl_eh}) and (\ref{new_expl_em})
can be re-expressed as \cite{racz_constraints} 
(see also  \cite{racz_geom_det,racz_geom_cauchy,racz_tdfd})
\begin{align} 
\mycal{L}_{\widehat n}({\rm\bf K}^l{}_{l}) - \widehat D^l {\rm\bf k}_{l} 
+ 2\,\dot{\widehat n}{}^l\, {\rm\bf k}_{l}
- [\,\boldsymbol\kappa-\tfrac12\, ({\rm\bf K}^l{}_{l})\,]\,
({\widehat K^{l}}{}_{l})  + \interior{\rm\bf K}{}_{kl}{\widehat K}{}^{kl}  = {}& 0 \,,   
\label{constr_mom2}  \\
\mycal{L}_{\widehat n} {\rm\bf k}{}_{i}  
+ ({\rm\bf K}^l{}_{l})^{-1}[\,\boldsymbol\kappa\,\widehat D_i ({\rm\bf K}^l{}_{l})
-2\, {\rm\bf k}{}^{l}\widehat D_i{\rm\bf k}{}_{l}\,] 
+ (2\,{\rm\bf K}^l{}_{l})^{-1}\widehat D_i\boldsymbol\kappa_0 {}& \nonumber \\ 
+ ({\widehat K^{l}}{}_{l})\,{\rm\bf k}{}_{i}
+ [\,\boldsymbol\kappa-\tfrac12\, ({\rm\bf K}^l{}_{l})\,]\,\dot{\widehat n}{}_i  
- \dot{\widehat n}{}^l\,\interior{\rm\bf K}_{li}  
+ \widehat D^l \interior{\rm\bf K}{}_{li}  = {}& 0 \, , \label{constr_mom1}
\end{align}
where $\boldsymbol\kappa$ and $\boldsymbol\kappa_0$ are given by the algebraic expressions 
\begin{equation} \label{constr_ham_n} 
\boldsymbol\kappa= (2\,{\rm\bf K}^l{}_{l})^{-1}[\, 2\,{\rm\bf k}{}^{l}{\rm\bf k}{}_{l} - \tfrac12\,({\rm\bf K}^l{}_{l})^2 - \boldsymbol\kappa_0 \,] \,,
\end{equation}
\begin{equation} \label{constr_ham_n0} 
\boldsymbol\kappa_0= {}^{{}^{(3)}}\hskip-1mm R - \interior{\rm\bf K}{}_{kl}\,\interior{\rm\bf K}{}^{kl}\,,
\end{equation}
and where $\widehat D_i$ and $\widehat R$ denote the covariant derivative operator and scalar curvature associated with
$\widehat \gamma_{ij}$, respectively, and $\dot{\widehat n}{}_k={\widehat n}{}^lD_l{\widehat n}{}_k=-{\widehat D}_k(\ln{\widehat N})$.

\medskip

Note that (\ref{constr_ham_n}) replaces the Hamiltonian constraint (\ref{new_expl_eh}) which acquires, 
thereby, an algebraic form (for more details see \cite{racz_constraints}).
Note also that in virtue of (\ref{constr_mom2})-(\ref{constr_ham_n0}) the four basic variables
$\boldsymbol\kappa, {\rm\bf k}{}_{i}, {\rm\bf K}^l{}_{l}$ are subject to the constraints whereas the remaining eight
varibales, represented by the fields 
$\widehat N,\widehat N^i,\widehat \gamma_{ij}, \interior{\rm\bf K}_{ij}$, are 
freely specifiable throughout $\Sigma$.

\section{The Newman-Penrose $\eth$ and $\,\overline\eth$ operators}
\label{ethBAReth}
\setcounter{equation}{0}

Equations (\ref{constr_mom2})-(\ref{constr_ham_n}) are intended to be solved by decomposing the involved 
basic variables in terms of spin-weighted spherical fields. In doing so we shall replace
all angular derivatives by the Newman-Penrose $\eth$ and
$\,\overline\eth$ operators  \cite{newman_penrose,goldberg_et_al},
using the notation introduced in \cite{jeff_edth,jeff_edth_2} throughout this paper.

\medspace

Consider first the unit sphere metric $q_{ab}$, given in standard $(\theta,\phi)$ coordinates by 
\begin{equation}\label{le}
ds^2= q_{ab}\, \d x^a \d x^b = \d\theta^2 + \sin^2\theta\, \d\phi^2\,.
\end{equation}

In terms of the complex stereographic coordinate\,\footnote{Expressions relevant for the south hemisphere will
	only be given explicitly. From these the ones which apply to the north hemisphere can be deduced
	by using the replacement $z_N=1/z_S$ \cite{jeff_edth,jeff_edth_2}.}
\begin{equation}
z = e^{-i\,\phi}\cot\frac{\theta}2=z_1+\mathbbm{i}\,z_2\,,
\end{equation}
on the unit sphere $\mathbb{S}^2$, the line element (\ref{le}) can also be written as  
\begin{equation}\label{conf_flat}
ds^2 = 4\,(1+ z \,\overline z)^{-2}\left[\,(\d z_1)^2+(\d z_2)^2\,\right]\,.
\end{equation}

Choose now the complex dyad on $\mathbb{S}^2$
\begin{equation}\label{dyad}
q^a = {2^{-1}}{P}\left[\,(\partial_{z_1})^a + \mathbbm{i}\,(\partial_{z_2})^a\right] = P\left(\partial_{\,\overline z}\right)^a\,,
\end{equation}
where 
\begin{equation}
P = 1+z\,\overline z \,.
\end{equation}
We also have
\begin{equation}\label{dual}
q_a = q_{ab}\,q^b = 2\, P^{-1}\left[\,(\d z_1)_a + \mathbbm{i}\,(\d z_2)_a\right]= 2\,P^{-1}\left(\,\d z\right)_a\,.
\end{equation}

Note that the complex dyad $q^a$ has normalization
\begin{equation}\label{unit_metr-norm}
q^a \,\overline q_a =2\,, \quad q^a q_a =0\,, 
\end{equation}
and that the unit sphere metric $q_{ab}$ satisfies
\begin{equation}\label{unit_metr}
q _{ab} = q_{(a} \,\overline q_{b)}\,,  
\quad q^{ab} = q^{(a} \,\overline q{}^{\,b)}\,, \quad q^{ae} q_{eb} =\delta^a{}_b\,.
\end{equation}
Note also that the metric (\ref{conf_flat}) is conformally flat,
\begin{equation}\label{conf_flat2}
q_{ab}= \Omega^2\, \delta_{ab} \, ,
\end{equation}
with conformal factor 
\begin{equation}\label{conf_factor}
\Omega={2}\,{(1+z \,\overline z)^{-1}} = 2\, P^{-1}\,.
\end{equation}

\medskip

The Newman-Penrose $\eth$ and $\,\overline\eth$ operators are then given by (see, e.g.~(A4) in \cite{jeff_edth})
\begin{align}
\eth\,\mathbb{L} = {}& P^{1-s}\,\partial_{\,\overline z} \left( P^s\,\mathbb{L} \right) \label{eth-def1} \\
\,\overline\eth\,\mathbb{L} = {}& P^{1+s}\,\partial_{z} \left( P^{-s}\,\mathbb{L} \right) \,,\label{eth-def2}
\end{align}
where the spin-weight $s$ function $\mathbb{L}$ on the unit two-sphere is defined by the contraction 
\begin{equation}\label{eth-def3}
\mathbb{L}=q^{a_1}\dots q^{a_s}\, \mathbf{L}_{({a_1}\dots{a_s})} 
\end{equation}
for some totally symmetric traceless tensor field $\mathbf{L}_{{a_1}\dots{a_s}}$ on $\mathbb{S}^2$. 

\medskip

As pointed out in \cite{jeff_edth,jeff_edth_2}, this choice of $\eth$ and $\,\overline\eth$ corresponds to the
standard conventions in \cite{newman_penrose, goldberg_et_al,jeff_edth_2}. Therefore, the action of $\eth$ and
$\,\overline\eth$ on spin-weighted spherical harmonics ${}_{s}\mathbb{Y}_{\,l,m}$ is given by  (see e.g.~(2.6)-(2.8) in \cite{goldberg_et_al})
\begin{align}
{}_{s}\,\overline{\mathbb{Y}}_{\,l,m}= {}& (-1)^{m+s}{}_{-s}\mathbb{Y}_{\,l,m} \\
\eth\,{}_{s}\mathbb{Y}_{\,l,m}= {}& \sqrt{(l-s)(l+s+1)}\,{}_{s+1}\mathbb{Y}_{\,l,m} \\
\,\overline\eth\,{}_{s}\mathbb{Y}_{\,l,m}= {}& -\sqrt{(l+s)(l-s+1)}\,{}_{s-1}\mathbb{Y}_{\,l,m}\\
\,\overline\eth\,\eth\,{}_{s}\mathbb{Y}_{\,l,m}= {}& -(l-s)(l+s+1)\,{}_{s}\mathbb{Y}_{\,l,m}\,.
\end{align}

\medskip

Also, the $\eth$ and $\,\overline\eth$ operators are related to the torsion free covariant derivative operator $\mathbb{D}_a$
determined by $q_{ab}$ by (see the Appendix of this paper for verification)
\begin{align}
\eth\,\mathbb{L} = {}& q^b q^{a_1}\dots q^{a_s}\, \mathbb{D}_b\mathbf{L}_{({a_1}\dots{a_s})}\,,\\
\,\overline\eth\,\mathbb{L} = {}& {\,\overline q}^b q^{a_1}\dots q^{a_s}\, \mathbb{D}_b\mathbf{L}_{({a_1}\dots{a_s})}\,.
\end{align}

\medskip

The applied conventions are such that the volume element $\boldsymbol\epsilon_{AB}$  on $\mathbb{S}^2$ is
$\boldsymbol\epsilon_{AB}=i\,q_{[A} \,\overline q_{B]}$ and for a spin-weight $s$ field $\mathbbm{f}$ the relation
$\left[\,\overline\eth, \eth\,\right]\,\mathbbm{f} = 2\,s\,\mathbbm{f}$ holds on $\mathbb{S}^2$.

\section{Reduction to spin-weighted fields}\label{spinwv}
\setcounter{equation}{0}

Consider now one of the level surfaces $\mycal{S}_{\rho_0}$ of the foliation $\mycal{S}_\rho$. As $\mycal{S}_{\rho_0}$ is
diffeomorphic to the unite sphere $\mathbb{S}^2$ we may assume that standard spherical coordinates $(\theta,\phi)$ as
in (\ref{le}) are chosen on  $\mycal{S}_{\rho_0}$.  
By using the vector field $\rho^i=\left(\partial_\rho\right)^i$ these coordinates can be Lie dragged onto
all the other leaves of the foliation $\mycal{S}_\rho$ by keeping their values constant along the integral curves of $\rho^i$.

\medskip

Note that in terms of the coordinates $(\theta,\phi)$, by making use of the line element (\ref{le}), the metric $q_{ab}$ can immediately be defined on each of the level surfaces $\mycal{S}_\rho$. Similarly, the complex dyad vector $q^a$ can be defined on the individual $\mycal{S}_\rho$ surfaces. 

\medskip

It is then an important consequence of the above construction that not only the coordinates $\theta$ and $\phi$ but  the complex dyad $q^a$, along with $q_a$ and the unit sphere metric $q_{ab}$, will be Lie dragged from $\mycal{S}_{\rho_0}$ onto the other level surfaces of the foliation $\mycal{S}_\rho$, i.e. 
\begin{equation}\label{lie_dragged}
\mycal{L}_\rho\, q^a=0\,, \quad  \mycal{L}_\rho\, q_a=0 \quad {\rm  and} \quad \mycal{L}_\rho\, q_{ab}=0  \,.
\end{equation}

\subsection{The decomposition of the metric $\widehat\gamma_{ab}$}

The metric $\widehat\gamma_{ab}$ induced on the $\mycal{S}_\rho$ level surfaces can then be decomposed as
\begin{equation}\label{ind_metr}
\widehat\gamma_{ab}=\mathbbm{a}\, q_{ab}+\interior\gamma_{ab}\,, 
\end{equation}
where 
\begin{equation}
\mathbbm{a}=\tfrac12\,\widehat\gamma_{ab}\,q^a \,\overline q^b
\end{equation}
is a positive spin-weight zero function on the $\mycal{S}_\rho$ level surfaces and
$\interior\gamma_{ab}$ is the trace-free part of $\widehat\gamma_{ab}$ with respect to the unit sphere metric $q_{ab}$, i.e.~
\begin{equation}\label{ind_metr_trf}
\interior\gamma_{ab}=\left[\delta_a{}^e\delta_b{}^f-\tfrac12\,q_{ab}\,q^{ef}\right]\widehat\gamma_{ef}=\widehat\gamma_{ab}-\mathbbm{a}\, q_{ab}\,.
\end{equation}

As $\interior\gamma_{ab}$ is a symmetric trace-free tensor it is given by
\begin{equation}\label{ind_metr_int}
\interior\gamma_{ab}=\tfrac12\left[ \mathbbm{b} \,\overline q_a \,\overline q_b + \,\overline{\mathbbm{b}}\, q_a q_b \right] \,,
\end{equation}
where the spin-weight $2$ function $\mathbbm{b}$ is given by the contraction
\begin{equation}
\mathbbm{b}=\tfrac12\,\widehat\gamma_{ab}\,q^a q^b=\tfrac12\,\interior\gamma_{ab}\,q^a q^b\,.
\end{equation}

It may also be verified that the inverse $\widehat\gamma^{ab}$ metric can be given by
\begin{equation}\label{inv_ind_metr}
\widehat\gamma^{ab}=\mathbbm{d}^{-1}\left\{\mathbbm{a}\, q^{ab}-\tfrac12\left[ \mathbbm{b} \,\overline q^a \,\overline q^b
+ \,\overline{\mathbbm{b}}\, q^a q^b \right]\right\}\,, 
\end{equation}
where 
\begin{equation}
\mathbbm{d}=\mathbbm{a}^2-\mathbbm{b}\,\overline{\mathbbm{b}}
\end{equation}
stands for the ratio  $\det(\widehat\gamma_{ab})/\det(q_{ab})$ of the determinants of
$\widehat\gamma_{ab}$ and $q_{ab}$. 

\medskip

As an immediate application, using (\ref{unit_metr}), (\ref{ind_metr}) and (\ref{ind_metr_trf}), along with the notation
\begin{equation}
{\mathbbm{k}}=q^l\,{\rm\bf k}{}_{l}\,,\quad \,\overline{\mathbbm{k}}=\,\overline q^l\,{\rm\bf k}{}_{l}\,,
\end{equation} 
${\rm\bf k}{}^{l} {\rm\bf k}{}_{l}$ can be expressed as 
\begin{align}
{\rm\bf k}{}^{l} {\rm\bf k}{}_{l}= {}& \widehat\gamma^{kl}{\rm\bf k}{}_{k}{\rm\bf k}{}_{l}=\tfrac12\,
\mathbbm{d}^{-1}\left\{\mathbbm{a} \left( q^k\,\overline q^l + q^l\,\overline q^k\right) - \left[\,\mathbbm{b}\, \,\overline q^k\,\overline q^l 
+ \,\overline{\mathbbm{b}}\, q^l q^k\, \right] \right\}{\rm\bf k}{}_{k}{\rm\bf k}{}_{l} \nonumber \\ 
= {}& \tfrac12\,\,\mathbbm{d}^{-1} [\, 2\,\mathbbm{a}\,{\mathbbm{k}}\,\overline{\mathbbm{k}}
- \mathbbm{b}\,\overline{\mathbbm{k}}^2- \,\overline{\mathbbm{b}}\,\mathbbm{k}^2\,]\,.
\end{align}

\subsection{Terms involving the covariant derivative $\widehat D_{A}$}

The covariant derivative operators $\widehat D_{A}$ and ${\mathbb D}_{A}$ can be related by the (1,2) type tensor field
(see e.g.~(3.1.28) and (D.3) in \cite{wald})
\begin{align}
{C^e}{}_{ab}=  \tfrac12\,\widehat\gamma^{ef}\left\{ {\mathbb D}_{a}\widehat\gamma_{fb}
+ {\mathbb D}_{b}\widehat\gamma_{af}-{\mathbb D}_{f}\widehat\gamma_{ab}\right\} \,. 
\end{align} 
In particular,
\begin{equation}
\widehat D_{a} {\rm\bf k}{}_{b}={\mathbb D}_{a} {\rm\bf k}{}_{b}-{C^e}{}_{ab}  {\rm\bf k}{}_{e} \,  ,
\end{equation}
and thereby 
\begin{align}\label{divk}
             \widehat D^{l} \,{\rm\bf k}{}_{l} = {}& \widehat\gamma^{kl} \,\widehat D_{k} \,{\rm\bf k}{}_{l}
             =\tfrac12\,\mathbbm{d}^{-1}\left\{\mathbbm{a} \left( q^k\,\overline q^l + q^l\,\overline q^k\right) 
             - \left[\,\mathbbm{b}\, \,\overline q^k\,\overline q^l + \,\overline{\mathbbm{b}}\, q^l q^k\, \right] \right\} \widehat D_{k} \,{\rm\bf k}{}_{l} \nonumber \\
            = {}& \tfrac14\,\mathbbm{d}^{-1}\left\{ 2 \mathbbm{a}
             \left(\,\eth\,\overline{\mathbbm{k}}-\mathbb{B}\,\overline{\mathbbm{k}}  \right) 
             - \mathbbm{b} \left(2\,\overline{\eth}\,\overline{\mathbbm{k}}-\,\overline{\mathbb{C}}\,\mathbbm{k}-\,\overline{\mathbb{A}}\,\overline{\mathbbm{k}}  \right) 
             + ``\,CC\,"\right\} \,,
\end{align}
where 
\begin{align} \label{ABC}
\mathbb{A} = {}&  q^a q^b {C^e}{}_{ab}\,\overline q_e = \mathbbm{d}^{-1}\left\{ \mathbbm{a}\left[2\,\eth\,\mathbbm{a} -\,\overline{\eth}\,\mathbbm{b}\right] 
   -  \,\overline{\mathbbm{b}}\,\eth\,\mathbbm{b} \right\} \nonumber \\
\mathbb{B} = {}&  \,\overline q^a q^b {C^e}{}_{ab}\,q_e = \mathbbm{d}^{-1}\left\{ \mathbbm{a}\,\overline{\eth}\,\mathbbm{b}
  - \mathbbm{b}  \,\eth\,\overline{\mathbbm{b}}\right\} \\
\mathbb{C} = {}&  q^a q^b {C^e}{}_{ab}\,q_e = \mathbbm{d}^{-1}\left\{ \mathbbm{a}\,\eth\,\mathbbm{b} -  \mathbbm{b}\left[2\,\eth\,\mathbbm{a}
   -\,\overline{\eth}\,\mathbbm{b}\right] \right\} \,. \nonumber
\end{align}
Hereafter $``\,CC\,"$ stands for the complex conjugate of the terms at the pertinent level of the hierarchy. 

\medskip

The relation 
\begin{align}\label{2kDk}
\hskip-1cm
     [\,2\,{\rm\bf k}{}^{l}\widehat D_{i} \,{\rm\bf k}{}_{l}\,]\,q^i= & [\,2\,\widehat\gamma^{kl}{\rm\bf k}{}_{k}\,{\widehat D}_{i} \,{\rm\bf k}{}_{l} \,]\,q^i  \\ 
     = {}& [\,\mathbbm{d}^{-1}\left\{\mathbbm{a} \left( q^k\,\overline q^l + q^l\,\overline q^k\right) - \left[\,\mathbbm{b}\, \,\overline q^k\,\overline q^l 
     + \,\overline{\mathbbm{b}}\, q^l q^k\, \right] \right\} {\rm\bf k}{}_{k}\,{\widehat D}_{i} \,{\rm\bf k}{}_{l}\, ]\,q^i \nonumber\\ 
     =  \tfrac12\,\,\mathbbm{d}^{-1}\left\{ \left(\mathbbm{a} \,\mathbbm{k}-\mathbbm{b}\,\overline{\mathbbm{k}}\right)\right.{}
     &\hskip-2mm\left. \left[2\,\eth\,\overline{\mathbbm{k}}-\,\overline{\mathbb{B}}\,\mathbbm{k}-\mathbb{B}\,\overline{\mathbbm{k}}  \right] 
     + \left(\mathbbm{a} \,\overline{\mathbbm{k}}-\,\overline{\mathbbm{b}}\,\mathbbm{k}\right)\left[2\,\eth\,\mathbbm{k}
     -\mathbb{C}\,\overline{\mathbbm{k}}-\mathbb{A}\,\mathbbm{k}  \right] \right\} \,, \nonumber
\end{align}
can also be verified.

\subsection{The scalar curvature ${}^{{}^{(3)}}\hskip-1mm R$}

In expressing the scalar curvature  ${}^{{}^{(3)}}\hskip-1mm R$ in terms of spin-weighted fields
we can use the relation
\begin{equation}\label{R3}
{}^{{}^{(3)}}\hskip-1mm R= \widehat R - [\,2\,\mycal{L}_{\widehat n} ({\widehat K^l}{}_{l}) + ({\widehat K^{l}}{}_{l})^2 
+ \widehat K_{kl} \widehat K^{kl} + 2\,{\widehat N}^{-1}\,\widehat D^l \widehat D_l \widehat N \,]\,,
\end{equation}
were the scalar curvature $\widehat R$ of the metric $\widehat\gamma_{ab}$ is given by
\begin{align}
\widehat R= \tfrac12\,\mathbbm{d}^{-1} {}& \left\{  2\,\mathbbm{a} -\eth\,\overline{\eth}\,\mathbbm{a}+ \,\overline{\eth}^2\,\mathbbm{b}
  + \tfrac12\,\mathbbm{d}^{-1}\left[ 2\,\left(\eth\,\mathbbm{a}\right)\left(\mathbbm{a}\,\overline{\eth}\,\mathbbm{a}
  -\mathbbm{a}\,\eth\,\overline{\mathbbm{b}}-\mathbbm{b}\,\overline{\eth}\,\overline{\mathbbm{b}}\right) \right.\right.  \\ 
& \left.\left. \hskip8mm + \left(\eth\,\mathbbm{b}\right)\left(\,\overline{\mathbbm{b}}\,\eth\,\overline{\mathbbm{b}}
   +\tfrac12\,\mathbbm{a}\,\overline{\eth}\,\overline{\mathbbm{b}}\right) + \left(\eth\,\overline{\mathbbm{b}}\right)\left(\mathbbm{b}\,\eth\,\overline{\mathbbm{b}}
  -\tfrac12\,\mathbbm{a}\,\overline{\eth}\,\mathbbm{b}\right)\right]\right\} + ``\,CC\,"\,,\nonumber
\end{align}
or, by using (\ref{ABC}) to replace first order derivatives, $\widehat R$ can also be given in the shorter form   
\begin{equation}\label{R2}
\widehat R = \,\widehat{\mathbb{R}} =  {\mathbbm{b}}^{-1}\left\{ \,\overline{\eth}\,\mathbb{C} -  \eth\,\mathbb{B} 
+ \tfrac12\,\left[ \mathbb{C}\,\overline{\mathbb{A}} + \mathbb{A}\,\mathbb{B} - \mathbb{B}^2 - \,\overline{\mathbb{B}}\,\mathbb{C} \right] \right\} \,.
\end{equation}

\subsection{Terms involving the lapse $\widehat{N}$}

Using the notation 
\begin{equation}
\,\widehat{\mathbb{N}}={\widehat N}\,
\end{equation}
we obtain
\begin{align}
{}& \widehat D^l\widehat D_l {\widehat N} = \widehat \gamma^{kl} [\,\widehat D_k\mathbb{D}_l {\widehat N}\,]  
  = \widehat \gamma^{kl} [\,\mathbb{D}_k\mathbb{D}_l {\widehat N} - C^{f}{}_{kl} \mathbb{D}_f {\widehat N}\,] 
\\ {}&  \phantom{\widehat D^l\widehat D_l {\widehat N}}  
=  \mathbbm{d}^{-1}\left\{\mathbbm{a}\, q^{kl}-\tfrac12\left[ \mathbbm{b} \,\overline q^k \,\overline q^l 
+ \,\overline{\mathbbm{b}}\, q^k q^l \right]\right\} [\,\mathbb{D}_k\mathbb{D}_l {\widehat N} 
- \tfrac12\,C^{f}{}_{kl}\left[q_f\,\overline q^e+\,\overline q_f q^e\right] \mathbb{D}_e {\widehat N}\,] \nonumber  \\ {}&  
= \tfrac12\,{\mathbbm{d}}^{-1}[\,\mathbbm{a} \{\,(\eth\,\overline{\eth}\,\widehat{\mathbb{N}}) 
   - \mathbb{B}\,(\,\overline{\eth}\,\widehat{\mathbb{N}}) \,\}   - \mathbbm{b} \,\{\,(\,\overline{\eth}^2\,\widehat{\mathbb{N}}) 
   - \tfrac12\,\overline{\mathbb{A}}\,(\,\overline{\eth}\,\widehat{\mathbb{N}}) 
   - \tfrac12\,\overline{\mathbb{C}} \, ({\eth}\,\widehat{\mathbb{N}}) \,\} + ``\,CC\,"  \,] \,. \nonumber
\end{align}
In virtue of the relation $\dot{\widehat n}{}_k={\widehat n}{}^lD_l{\widehat n}{}_k=-{\widehat D}_k(\ln{\widehat N})$ we also have 
\begin{equation}
      q^{i\,} \dot{\widehat n}{}_i = -  \widehat{\mathbb{N}}^{-1}  \eth \widehat{\mathbb{N}}
\end{equation}
and 
\begin{equation}\label{ndotk}
  \mathbbm{k}^{i\,} \dot{\widehat n}{}_i = - (2\,\mathbbm{d} \,\widehat{\mathbb{N}})^{-1}   \{
     \,(\eth\,\widehat{\mathbb{N}}) \,[\, \mathbbm{a}\,\overline {\mathbbm{k} }-
    \overline {\mathbbm{b}} \, \mathbbm{k} \,]
   + ``\,CC\,"  \, \}\,.
\end{equation}

\subsection{Terms involving the shift $\widehat N^i$ and ${\rm\bf K}^l{}_{l}$}

By making use of the relations 
\begin{equation}
\widetilde{\mathbb{N}}=q_i\widehat N^i= q_i \widehat\gamma{}^{ij}  \widehat N_j 
= \mathbbm{d}^{-1} (\mathbbm{a}\,q^j - \mathbbm{b}\,\overline q^j)\,{\widehat N}{}_{j}
=  \mathbbm{d}^{-1} (\mathbbm{a}\,\mathbb{N} - \mathbbm{b}\,\overline{\mathbb{N}})
\end{equation}
or alternatively 
\begin{equation}
\mathbb{N}=q^l {\widehat N}{}_{l}=q^l \widehat\gamma{}_{lk} {\widehat N}{}^{k} = (\mathbbm{a}\,q_k + \mathbbm{b}\,\overline q_k)\,{\widehat N}{}^{k}
= \mathbbm{a}\,\widetilde{\mathbb{N}} + \mathbbm{b}\,\overline {\widetilde{\mathbb{N}}} \,, 
\end{equation}
the Lie derivative $\mycal{L}_{\widehat n}\,({\rm\bf K}^l{}_{l})$ appearing in (\ref{constr_mom2}) can be expressed as 
\begin{align}\label{Lie_K} 
          \mycal{L}_{\widehat n}\,({\rm\bf K}^l{}_{l}) 
          = {}& {\widehat n}{}^i D_i{\rm\bf K}^l{}_{l} = {{\widehat N}^{-1}} [\, (\partial_\rho)^i -\widehat N^i \,]\, D_i {\rm\bf K}^l{}_{l} 
          = {\widehat N}^{-1} [\, \partial_\rho{\rm\bf K}^l{}_{l}  -\widehat N^i \, \mathbb{D}_i{\rm\bf K}^l{}_{l} \,]\nonumber  \\ 
          = {}&  \mycal{L}_{\widehat n}\,\mathbb{K} = \,\widehat{\mathbb{N}}^{-1}[\, (\partial_\rho \mathbb{K})
           -\tfrac12\, \widetilde{\mathbb{N}} \,(\,\overline{\eth}\, \mathbb{K}) 
           - \tfrac12\,\overline{\widetilde{\mathbb{N}}}\, (\eth\,\mathbb{K})    \,]\,,
\end{align}
where
\begin{equation}
\mathbb{K} = {\rm\bf K}^l{}_{l}= \widehat\gamma^{kl} \,{\rm\bf K}{}_{kl}
\end{equation} 
and we have used  $\widehat N^i D_i{\rm\bf K}^l{}_{l} = \widehat N^i\mathbb{D}_i{\rm\bf K}^l{}_{l}
=\tfrac12\,\widehat N^i\left(q_i\,\overline q^j+\,\overline q_iq^j\right) \mathbb{D}_j{\rm\bf K}^l{}_{l}$\, .

\subsection{Terms involving the trace-free part of ${\rm\bf K}{}_{kl}$}

By setting 
\begin{equation}
            \interior{\mathbb{K}} = q^kq^l\,\interior{\rm\bf K}{}_{kl}
\end{equation}
and
\begin{equation}
         \inbullet{\mathbb{K}} = q^{k}\,\overline q^l\,\interior{\rm\bf K}{}_{kl}\,,
\end{equation}
in virtue of  (\ref{intK}), we obtain
\begin{equation}\label{decomp_bfK}
        \interior{\rm\bf K}{}_{ij}=\tfrac12\,q{}_{ij}\,\inbullet{\mathbb{K}}+\tfrac14\,[\, q_{i}q_j\,\overline{\interior{\mathbb{K}}}
         + \overline q_{i}\overline q_j\,\interior{\mathbb{K}}\,] \,.
\end{equation}

\medskip

Note that, since $\interior{\rm\bf K}{}_{kl}$ is trace free,  $\inbullet{\mathbb{K}}$ and $\interior{\mathbb{K}}$ are not functionally independent.
Indeed, the trace-free condition $\widehat\gamma^{kl}\,\interior{\rm\bf K}{}_{kl}=0$ implies 
\begin{equation}\label{relation}
            \inbullet{\mathbb{K}}= (2\,\mathbbm{a})^{-1} [\,\mathbbm{b}\,\overline{\interior{\mathbb{K}}} +  \overline{\mathbbm{b}}\,\interior{\mathbb{K}} \,]\,.
\end{equation}

For both $\mathbbm{a}^{-1}$ and $\inbullet{\mathbb{K}}$, to be well-defined $\mathbbm{a}$ cannot vanish.
This is, however, guaranteed because $\widehat{\gamma}_{ij}$ is a positive definite Riemannian metric so
that $\mathbbm{d}=\mathbbm{a}^2-\mathbbm{b}\,\overline{\mathbbm{b}}$ is positive. 

\medskip

We then have
\begin{equation}\label{qndotK}
           q^{i\,} \dot{\widehat n}{}^{k\,} \interior{\rm\bf K}{}_{ki}	 =
	   -\tfrac12 ( {\widehat{\mathbb{N}}}\,{\mathbbm{d}})^{-1} \left[
	      \mathbbm{a}\,(\overline \eth\, \widehat{\mathbb{N}} ) \, \interior{\mathbb{K}}    
	      +  \mathbbm{a}\,( \eth {\widehat{\mathbb{N}}} ) \, \inbullet{\mathbb{K}}  
	      -\mathbbm{b}\,(\overline \eth {\widehat{\mathbb{N}}} ) \, \inbullet{\mathbb{K}} 
	      - \overline{\mathbbm{b}}\,(\eth {\widehat{\mathbb{N}}} ) \,
	      \interior{\mathbb{K}}   \right]
\end{equation}
\begin{align}\label{qDK}
           q^i \widehat D^{k\,} \interior{\rm\bf K}{}_{ki}	 &=
	   \tfrac12\,{\mathbbm{d}}^{-1} \left\{
	      \mathbbm{a}\,\overline \eth \, \interior{\mathbb{K}}    
	      +  \mathbbm{a}\,\eth \, \inbullet{\mathbb{K}}  
	      -\mathbbm{b} \,\overline \eth \, \inbullet{\mathbb{K}} 
	      - \overline{\mathbbm{b}}\, \eth \, \interior{\mathbb{K}}   \right\}
	      \nonumber \\
	& - \frac{ \mathbbm{a}} {4\mathbbm{d}} \left\{
	       3\,\overline{\mathbbm{B}} \,\interior{\mathbb{K}} +3\,{\mathbbm{B}}\, \inbullet{\mathbb{K}} 
	         +{\mathbbm{A}} \,\inbullet{\mathbb{K}} +{\mathbbm{C}}\,
		 \overline{\interior{\mathbb{K}}} \right\} 
		 \nonumber \\
     & + \frac{ \mathbbm{b}} {4\mathbbm{d}} \left\{
	       \overline{\mathbbm{C}} \,\interior{\mathbb{K}} +\overline{\mathbbm{A}} \,\inbullet{\mathbb{K}} 
	         +\overline{\mathbbm{B}} \,\inbullet{\mathbb{K}} +{\mathbbm{B}} \,\overline{\interior{\mathbb{K}}} \right\} 
	+ \frac{\overline{ \mathbbm{b}}} {2\mathbbm{d}} \left\{
	     \mathbbm{A}\,\interior{\mathbb{K}}  +\mathbbm{C}\, \inbullet{\mathbb{K}}  \right\}  
\end{align}
\begin{align}\label{tracefreebfcurv_sq}
            \interior{\rm\bf K}{}_{ij}   \interior{\rm\bf K}{}^{ij}  
	    = {} & \tfrac14  \,{\mathbbm{d}}^{-2} \left[
	       \left \{ \, {\overline { \interior{\mathbb{K}}}} \,
	         (\, {\mathbbm{a}}^2\,{ \interior{\mathbb{K}}}   
		+ {\mathbbm{b}}^2  {\,\overline { \interior{\mathbb{K}}}}  
		-4\,\mathbbm{a} \,\mathbbm{b} \,\inbullet{\mathbb{K}}\, )
		+ ``\,CC\," \right \} 
		+  2\,  ({\mathbbm{a}}^2+{\mathbbm{b}}\,\overline {\mathbbm{b}} )
		\,\inbullet{\mathbb{K}}^2\right] \,.
\end{align}

\subsection{The determination of $\mycal{L}_{\widehat n}\,{\rm\bf k}{}_{l}$ and $\mycal{L}_{\rho} {\rm\bf k}{}_{i}$}

The Lie derivative $\mycal{L}_{\widehat n}\,{\rm\bf k}{}_{l}$ appearing in (\ref{constr_mom1}), can be re-expressed as follows.

\medskip

Note first that  
\begin{equation}
\left(\mycal{L}_{\widehat n}\,{\rm\bf k}{}_{l}\right)  {\widehat n}{}^l= \mycal{L}_{\widehat n}\left({\rm\bf k}{}_{l}\,{\widehat n}{}^l\right) =0
\end{equation}
which implies
\begin{equation}
\mycal{L}_{\widehat n}\,{\rm\bf k}{}_{l}= \widehat\gamma_{\,l}{}^i \mycal{L}_{\widehat n}\,{\rm\bf k}{}_{i}\,.
\end{equation}
Then, it is straightforward to verify that 
\begin{align}
\mycal{L}_{\widehat n}\,{\rm\bf k}{}_{l} = {}& \widehat\gamma_{\,l}{}^i \mycal{L}_{\widehat n}\,{\rm\bf k}{}_{i}=
     {{\widehat N}^{-1}} \widehat\gamma_{\,l}{}^i\,[\, \mycal{L}_{\rho} {\rm\bf k}{}_{i} - \mycal{L}_{\widehat N} {\rm\bf k}{}_{i} \,] \nonumber \\ = {}&  
    {{\widehat N}^{-1}} [\, \widehat\gamma_{\,l}{}^i(\mycal{L}_{\rho} {\rm\bf k}{}_{i})  - \widehat N^f \widehat D_f {\rm\bf k}{}_{l}
     - {\rm\bf k}{}_{f} \widehat D_l \widehat N^f  \,] \nonumber \\ = {}&  
        {{\widehat N}^{-1}} [\, \widehat\gamma_{\,l}{}^i(\mycal{L}_{\rho} {\rm\bf k}{}_{i})  - \widehat N^f \mathbb D_f {\rm\bf k}{}_{l} 
        - {\rm\bf k}{}_{f} \mathbb D_l \widehat N^f  \,] \,,
\end{align}
where in the second line we have used the freedom in choosing a torsion free connection when evaluating $\mycal{L}_{\widehat N} {\rm\bf k}{}_{i}$ .

\medskip

In determining $q^l \mycal{L}_{\widehat n}\,{\rm\bf k}{}_{l} $ we use 
\begin{equation}
            q^l\widehat\gamma_{\,l}{}^i(\mycal{L}_{\rho} {\rm\bf k}{}_{i})=q^l q_{\,l}{}^i(\mycal{L}_{\rho} {\rm\bf k}{}_{i}) 
            = (\partial_\rho \mathbbm{k})
\end{equation}
and
\begin{align}
                q^l\,[\,\widehat N^f \mathbb D_f {\rm\bf k}{}_{l}\, + {\rm\bf k}{}_{f} \mathbb D_l \widehat N^f ] = 
		\tfrac{1}{2}[  \, {\widetilde{\mathbb{N}}} \, \,\overline{\eth}\,\mathbbm{k} 
		 + \,\overline {\widetilde{\mathbb{N}}} \, \eth\,\mathbbm{k}  \, ] 
		 +\tfrac{1}{2} [ \, \mathbbm{k} \, \eth \,\overline {\widetilde{\mathbb{N}}} 
		 + \,\overline {\mathbbm{k}} \, \eth {\widetilde{\mathbb{N}}}  \,  ] .
\end{align}
Then
\begin{equation}\label{liek}
           q^l \mycal{L}_{\widehat n}\,{\rm\bf k}{}_{l} ={\widehat{\mathbb{N}}^{-1}} \left( \partial_\rho \mathbbm{k}
	     - \tfrac12 [ \, {\widetilde{\mathbb{N}}} \, \,\overline{\eth}\,\mathbbm{k} 
		 + \,\overline {\widetilde{\mathbb{N}}} \, \eth\,\mathbbm{k}  
		 + \mathbbm{k} \, \eth \,\overline {\widetilde{\mathbb{N}}} 
		 + \,\overline {\mathbbm{k}} \, \eth \,{\widetilde{\mathbb{N}}}  \,  ] \right) .
\end{equation}

\subsection{The decomposition of $\widehat D_{k} {\widehat N}{}_{l}$}

We also need to evaluate the auxiliary expressions 
$q^kq^l\,(\widehat D_{k} {\widehat N}{}_{l})$ and $\,\overline q^kq^l\,(\widehat D_{k} {\widehat N}{}_{l})$. 
To do so notice first that 
\begin{equation}
\widehat D_{k} {\widehat N}{}_{l}={\mathbb D}_{k} {\widehat N}{}_{l} -{C^f}{}_{kl}  {\widehat N}{}_{f}
\end{equation}
from which one gets
\begin{align}
          q^kq^l\,(\widehat D_{k} {\widehat N}{}_{l}) = {}& q^kq^l\,({\mathbb D}_{k} {\widehat N}{}_{l}) 
          - q^kq^l\,{C^f}{}_{kl}\,[ \tfrac12\,(q_f\,\overline q^e + \,\overline q_fq^e) ]  {\widehat N}{}_{e} \nonumber \\ 
          = {}&  \eth\,\mathbb{N} -\tfrac12\, \mathbb{C}\,\overline{\mathbb{N}} - \tfrac12\,\mathbb{A}\,\mathbb{N}\,,
\end{align}
\begin{align}
       \,\overline q^kq^l\,(\widehat D_{k} {\widehat N}{}_{l}) = {}& \,\overline q^kq^l\,({\mathbb D}_{k} {\widehat N}{}_{l}) 
       - \,\overline q^kq^l\,{C^f}{}_{kl}\,[ \tfrac12\,(q_f\,\overline q^e + \,\overline q_f\,q^e) ]  {\widehat N}{}_{e} \nonumber \\ 
       = {}&  \overline\eth\,\mathbb{N} -\tfrac12\, \mathbb{B}\,\overline{\mathbb{N}} -\tfrac12\,\overline{\mathbb{B}}\,\mathbb{N} \,.
\end{align}

\subsection{Terms involving $\widehat K_{ij}$}

Before determining $q^l\,[\,\widehat\gamma{}^{ef} {\rm\bf k}{}_{e} \widehat K_{fl}\,]$ we need also to evaluate 
the extrinsic curvature $\widehat K_{ij}$ of $\mycal{S}_\rho$ as given by (\ref{hatextcurv}),
\begin{align}\label{hatextcurv2}
 \widehat K_{ij}= {}& \tfrac12\,\mycal{L}_{\widehat n} {\widehat \gamma}_{ij}=\tfrac12\,{\widehat N}^{-1}[\,\mycal{L}_{\rho}{\widehat \gamma}_{ij}
     - ( \widehat D_i\widehat N_j + \widehat D_j\widehat N_i )] \\ = {}& \tfrac12\,{\,\widehat{\mathbb{N}}}^{-1}[(\partial_\rho\mathbbm{a})\,q_{ij}
     +\tfrac12\,[\left(\partial_\rho\mathbbm{b}\right)\,\overline q_i\,\overline q_j + \left(\partial_\rho\,\overline{\mathbbm{b}}\right) q_i q_j] 
     - (\widehat D_i\widehat N_j + \widehat D_j\widehat N_i )]\,, \nonumber
\end{align}
where in the last step (\ref{lie_dragged}) was applied.  As a result, 
\begin{align}\label{hatextcurv_trace}
              {}& \,\widehat{\mathbb{K}} = \widehat K{}^{\,l}{}_{l} =   \widehat \gamma^{ij} \widehat K_{ij}
                  = \mathbbm{d}^{-1}\left\{\mathbbm{a}\, q^{ij}-\tfrac12\left[ \mathbbm{b} \,\overline q^i \,\overline q^j
		  + \,\overline{\mathbbm{b}}\, q^i q^j \right]\right\}\widehat K_{ij} 
                  = \tfrac12\,({\,\widehat{\mathbb{N}}\,\mathbbm{d}})^{-1} \times \nonumber  \\  
                  {}& \times \bigl[\mathbbm{a} \{(\partial_\rho\mathbbm{a}) - q^i \,\overline q^j\,[\,\widehat D_i\widehat N_j + \widehat D_j\widehat N_i\,] \} 
                  - \mathbbm{b}\{(\partial_\rho\,\overline{\mathbbm{b}}) - \,\overline q^i \,\overline q^j (\widehat D_i\widehat N_j)\}  \bigr] + ``\,CC\,"  \nonumber  \\ 
                  {}& = \tfrac12\,({\,\widehat{\mathbb{N}}\,\mathbbm{d}})^{-1}\left\{\mathbbm{a}\,[\,(\partial_\rho\mathbbm{a}) - (\,\overline\eth\,\mathbb{N}) 
                   + \,\overline{\mathbb{B}}\,\mathbb{N} \,] \right. \nonumber 
                    \\ {}& \hskip2.5cm \left.  - \mathbbm{b}\,[\,(\partial_\rho\,\overline{\mathbbm{b}}) - \,(\,\overline{\eth}\,\overline{\mathbb{N}}) 
                +\tfrac12\, \,\overline{\mathbb{C}}\,\mathbb{N} +\tfrac12\, \,\overline{\mathbb{A}}\,\overline{\mathbb{N}} \,]  \right\} + ``\,CC\," \, .
\end{align}

Set now 
\begin{equation}\label{hatextcurv_qq}
              \,\indiamond{\mathbb{K}} =  q^i q^j\widehat K_{ij} 
                 =\tfrac12\,{\,\widehat{\mathbb{N}}}^{-1}\left\{2\,\partial_\rho\mathbbm{b} - 2\,\eth\,\mathbb{N}
                   + {\mathbb{C}}\,\overline {\mathbb{N}} +\mathbb{A}\, {\mathbb{N}} \,  \right\} \,, 
\end{equation}
\begin{equation}\label{hatextcurv_qbq}
              \,\inblacklozenge{\mathbb{K}} =  q^i \,\overline q^j\widehat K_{ij} 
                =\tfrac12\,{\,\widehat{\mathbb{N}}}^{-1}\left\{2\,\partial_\rho\mathbbm{a}
               - \,\overline \eth\,\mathbb{N}  - \eth\,\overline{\mathbb{N}}
                   + {\mathbb{B}}\,\overline {\mathbb{N}} +\,\overline{\mathbb{B}}\,{\mathbb{N}} \,  \right\} \,. 
\end{equation}
Then, because the symmetric 2-tensor $\widehat K{}^{\,l}{}_{l}$ is determined by three real functions,
it follows that $\inblacklozenge{\mathbb{K}}$, $\indiamond{\mathbb{K}}$ and $\widehat{\mathbb{K}}$
are functionally dependent. In determining their algebraic relation it is advantageous to introduce the auxiliary variables
\begin{equation}\label{hatextcurv_qbq0}
              {}^\star\hskip-.051mm \inblacklozenge{\mathbb{K}} =  q^i \,\overline q^j\,[ \widehat K_{ij} -\tfrac12\,\widehat \gamma_{ij} \widehat K{}^{\,l}{}_{l} ]
              = \inblacklozenge{\mathbb{K}} - {\mathbbm{a}} \,\widehat{\mathbb{K}}  
\end{equation}
\begin{equation}\label{hatextcurv_qbq1}
              {}^\star\hskip-.051mm \indiamond{\mathbb{K}} =  q^i \, q^j\,[ \widehat K_{ij} -\tfrac12\,\widehat \gamma_{ij} \widehat K{}^{\,l}{}_{l} ]
              = \indiamond{\mathbb{K}} - {\mathbbm{b}} \,\widehat{\mathbb{K}}  \,.
\end{equation}
The analog of the trace relation (\ref{relation}) then gives
\begin{equation}\label{relation1}
            {}^\star\hskip-.051mm \inblacklozenge{\mathbb{K}}= (2\,\mathbbm{a})^{-1} [\,\mathbbm{b}\,\overline{{}^\star\hskip-.051mm \indiamond{\mathbb{K}}} +  \overline{\mathbbm{b}}\,{}^\star\hskip-.051mm \indiamond{\mathbb{K}}  \,]\,,
\end{equation}
from which it follows, in virtue of (\ref{hatextcurv_qbq0}) and (\ref{hatextcurv_qbq1}), 
\begin{equation}\label{relation2}
            \inblacklozenge{\mathbb{K}}={\mathbbm{a}}^{-1}\{\, \mathbbm{d}\cdot\widehat{\mathbb{K}} + \tfrac12 \,[\,\mathbbm{b}\,\overline{\indiamond{\mathbb{K}}} +  \overline{\mathbbm{b}}\,\indiamond{\mathbb{K}}  \,]\,\}\, .
\end{equation}

\medskip

Then, by making use of all the above $\widehat K^{ij}$ related variables, we obtain
\begin{equation}\label{hatextcurv_uqq}
            q_i q_j \widehat K^{ij} 
                ={\mathbbm{d}}^{-2}\,[\, {\mathbbm{a}}^2\,{\indiamond{\mathbb{K}}}  
		+ {\mathbbm{b}}^2 {\,\overline{\,\indiamond{\mathbb{K}}} } 
		-2\,\mathbbm{a}\, \mathbbm{b} \,\inblacklozenge{\mathbb{K}} \,  ] 
\end{equation}
and
\begin{equation}\label{hatextcurv_uqbq}
            q_i \,\overline q_j \widehat K^{ij} 
                ={\mathbbm{d}}^{-2}[\,
		({\mathbbm{a}}^2+{\mathbbm{b}}\,\overline {\mathbbm{b}} )\,{\inblacklozenge{\mathbb{K}} } 
		-\mathbbm{a} \,\overline{\mathbbm{b}} \,\indiamond{\mathbb{K}} 
		-\mathbbm{a} \,\mathbbm{b} \, {\,\overline{\indiamond{\mathbb{K}}}} \,  ] \,.
\end{equation}
These relations, along with (\ref{decomp_bfK}), imply
\begin{align}\label{hatextcurv_sq}
               \interior{\rm\bf K}{}_{ij} \widehat K^{ij}   = {}& \tfrac14\, {\mathbbm{d}}^{-2} 
	       \left[ \,2\, \inbullet{\mathbb{K}}\left( [\,{\mathbbm{a}}^2+{\mathbbm{b}}\,\overline {\mathbbm{b}})\,]\,{\inblacklozenge{\mathbb{K}} } - {\mathbbm{a}}\,[\,\overline{\mathbbm{b}}\,{\indiamond{\mathbb{K}}}  
		+ {\mathbbm{b}} {\,\overline{\indiamond{\mathbb{K}}} } \,]\,\right) \right.  \nonumber \\ 
           {}& \left.\hskip1.2cm + \left\{ \,{\overline{\interior{\mathbb{K}}}} 
	         \,[\, {\mathbbm{a}}^2\,{\indiamond{\mathbb{K}}}  
		+ {\mathbbm{b}}^2 {\,\overline{\indiamond{\mathbb{K}}} } 
		-2\,\mathbbm{a} \,\mathbbm{b} \,\inblacklozenge{\mathbb{K}} \,]
		+ ``\,CC\," \right\} \right] 
\end{align}
and
\begin{align}
               \widehat K_{ij} \widehat K^{ij}   = {}& \tfrac14\, {\mathbbm{d}}^{-2} 
	       \left \{ \,\overline{\,\indiamond{\mathbb{K}}}
	         \left [ {\mathbbm{a}}^2\,\indiamond{\mathbb{K}} 
		+ {\mathbbm{b}}^2 \,\overline{\,\indiamond{\mathbb{K}} } 
		-4\,\mathbbm{a} \,\mathbbm{b} \,\inblacklozenge{\mathbb{K}} \right ]
		+ ``\,CC\," \right \} 
		+  \tfrac12\, {\mathbbm{d}}^{-2} 
	 ({\mathbbm{a}}^2+{\mathbbm{b}}\,\overline {\mathbbm{b}} )\,\inblacklozenge{\mathbb{K}}^2  .\nonumber 
\end{align}

Finally, the analogue of (\ref{Lie_K}) is  
\begin{align}\label{Lie_hatK} 
            \mycal{L}_{\widehat n}\,({\widehat K}^l{}_{l}) = {}& \mycal{L}_{\widehat n}\,{\,\widehat{\mathbb{K}}} 
            =\,\widehat{\mathbb{N}}^{-1}\left[ (\partial_\rho \,\widehat{\mathbb{K}} ) -\tfrac12\, \widetilde{\mathbb{N}} \,(\,\overline{\eth}\, \,\widehat{\mathbb{K}})
            - \tfrac12\,\overline{\widetilde{\mathbb{N}}}\, (\eth \,\widehat{\mathbb{K}})    \right]\,.
\end{align}

\section{The constraints in terms of spin-weighted variables}
\setcounter{equation}{0}

This section presents the explicit form of the constraints in terms of the spin-weighted fields introduced in Section \ref{spinwv}.
To provide a clear outline of the analytic setup, these spin-weighted fields are collected in Table\,\ref{table:data}\,.
\begin{table}[h]
	\centering  \hskip-.15cm
	\begin{tabular}{|c|c|c|} 
		\hline notation &  definition  & \hskip-0.7cm$\phantom{\frac{\frac12}{A}_{B_D}}$ spin-weight \\ \hline \hline
		
		$\mathbbm{a}$ &  $\tfrac12\,q^i\,\overline q^j\,\widehat\gamma_{aj}$  & \hskip-0.7cm$\phantom{\frac{\frac12}{A}_{B_D}}$ $0$ \\  \hline 
		
		$\mathbbm{b}$ &  $\tfrac12\,q^i q^j\,\widehat\gamma_{ij}$  & \hskip-0.7cm$\phantom{\frac{\frac12}{A}_{B_D}}$ $2$ \\  \hline 
		
		$\mathbbm{d}$ &  $\mathbbm{a}^2-\mathbbm{b}\,\overline{\mathbbm{b}}$  & \hskip-0.7cm$\phantom{\frac{\frac12}{A}_{B_D}}$ $0$ \\  \hline 
		$\mathbbm{k}$ &  $q^i {\rm\bf k}{}_{i}$  & \hskip-0.7cm$\phantom{\frac{\frac12}{A}_{B_D}}$ $1$ \\  \hline 
		
		$\mathbb{A}$ &  $q^a q^b {C^e}{}_{ab}\,\overline q_e= \mathbbm{d}^{-1}\left\{ \mathbbm{a}\left[2\,\eth\,\mathbbm{a}
		 -\,\overline{\eth}\,\mathbbm{b}\right] -  \,\overline{\mathbbm{b}}\,\eth\,\mathbbm{b} \right\} $  & \hskip-0.7cm$\phantom{\frac{\frac12}{A}_{B_D}}$ $1$ \\  \hline 
		
		$\mathbb{B}$ &  $\,\overline q^a q^b {C^e}{}_{ab}\,q_e = \mathbbm{d}^{-1}\left\{ \mathbbm{a}\,\overline{\eth}\,\mathbbm{b}
		 - \mathbbm{b}  \,\eth\,\overline{\mathbbm{b}}\right\}$ & \hskip-0.7cm$\phantom{\frac{\frac12}{A}_{B_D}}$ $1$ \\  \hline 
		
		$\mathbb{C}$ &  $q^a q^b {C^e}{}_{ab}\,q_e = \mathbbm{d}^{-1}\left\{ \mathbbm{a}\,\eth\,\mathbbm{b} -  \mathbbm{b}\left[2\,\eth\,\mathbbm{a} 
		-\,\overline{\eth}\,\mathbbm{b}\right] \right\}$ & \hskip-0.7cm$\phantom{\frac{\frac12}{A}_{B_D}}$ $3$ \\  \hline 
		
		$\,\widehat{\mathbb{R}}$ &  $\widehat R = {\mathbbm{b}}^{-1}\left\{ \,\overline{\eth}\,\mathbb{C} -  \eth\,\mathbb{B} 
		+ \tfrac12\,\left[ \mathbb{C}\,\overline{\mathbb{A}} + \mathbb{A}\,\mathbb{B} - \mathbb{B}^2
		 - \,\overline{\mathbb{B}}\,\mathbb{C} \right] \right\}$ & \hskip-0.7cm$\phantom{\frac{\frac12}{A}_{B_D}}$ $0$ \\  \hline 
		
		$\,\widehat{\mathbb{N}}$ &  $\widehat N$  & \hskip-0.7cm$\phantom{\frac{\frac12}{A}_{B_D}}$ $0$ \\  \hline
		$\mathbb{N}$ &  $q^i\widehat N_i= q^i \widehat\gamma{}_{ij} {\widehat N}{}^{j}$  & \hskip-0.7cm$\phantom{\frac{\frac12}{A}_{B_D}}$ $1$ \\  \hline
		
		$\widetilde{\mathbb{N}}$ &  $q_i\widehat N^i = q_i \,\widehat\gamma{}^{ij} {\widehat N}{}_{j}
		=\mathbbm{d}^{-1} (\mathbbm{a}\,\mathbb{N} - \mathbbm{b}\,\overline{\mathbb{N}})$  & \hskip-0.7cm$\phantom{\frac{\frac12}{A}_{B_D}}$ $1$ \\  \hline
		
		$\mathbb{K}$ &  ${\rm\bf K}^l{}_{l}= \widehat\gamma^{kl} \,{\rm\bf K}{}_{kl}
		$  & \hskip-0.7cm$\phantom{\frac{\frac12}{A}_{B_D}}$ $0$ \\  \hline 
		
		$\interior{\mathbb{K}}$ &  $q^kq^l\,\interior{\rm\bf K}{}_{kl}
		$  & \hskip-0.7cm$\phantom{\frac{\frac12}{A}_{B_D}}$ $2$ \\  \hline 
		
		$\inbullet{\mathbb{K}}$ &  $q^k\,\overline{q}^l\,\interior{\rm\bf K}{}_{kl}= (2\,\mathbbm{a})^{-1} [\,\mathbbm{b}\,\overline{\interior{\mathbb{K}}} 
		+  \overline{\mathbbm{b}}\,\interior{\mathbb{K}} \,] 
		$  & \hskip-0.7cm$\phantom{\frac{\frac12}{A}_{B_D}}$ $0$ \\  \hline   
		
		$\,\widehat{\mathbb{K}}$ &  ${\widehat K}^l{}_{l} = \widehat\gamma^{ij} \widehat K_{ij} $  & \hskip-0.7cm$\phantom{\frac{\frac12}{A}_{B_D}}$ $0$ \\  \hline 
		
		$\indiamond{\mathbb{K}}$ &  $q^i q^j\widehat K_{ij} 
		= \tfrac12\,{\,\widehat{\mathbb{N}}}^{-1}\left\{2\,\partial_\rho\mathbbm{b} - 2\,\eth\,\mathbb{N}
		+ {\mathbb{C}}\,\overline {\mathbb{N}} +\mathbb{A} \,{\mathbb{N}} \,  \right\} $ & \hskip-0.7cm$\phantom{\frac{\frac12}{A}_{B_D}}$ $2$ \\  \hline 
		
		$\inblacklozenge{\mathbb{K}}$ &  $q^k\,\overline{q}^l\,\widehat{K} {}_{kl} = {\mathbbm{a}}^{-1}\{\,\mathbbm{d}\cdot\widehat{\mathbb{K}} 
		+ \tfrac12 \,[\,\mathbbm{b}\,\overline{\indiamond{\mathbb{K}}} +  \overline{\mathbbm{b}}\,\indiamond{\mathbb{K}}  \,]\,\}
		$  & \hskip-0.7cm$\phantom{\frac{\frac12}{A}_{B_D}}$ $0$ \\  \hline                    
		
	\end{tabular}
	\caption{\small The spin-weighted fields as they appear in various terms of (\ref{eq:eth_constr_mom2})--(\ref{eq:eth_constr_mom1}).}\label{table:data}
\end{table}

\medskip

By applying these fields and their relations, the constraint system comprised of (\ref{constr_mom2})--(\ref{constr_ham_n}) takes the form 
\begin{align} 
          \partial_\rho \mathbb{K} {}& - \tfrac12\,\widetilde{\mathbb{N}} \,(\,\overline{\eth}\, \mathbb{K}) -\tfrac12\, \,\overline{\widetilde{\mathbb{N}}}
            \, (\eth\,\mathbb{K}) - \tfrac12\,\widehat{\mathbb{N}}\,\mathbbm{d}^{-1}\left\{\, \mathbbm{a}\,(\eth\,\overline{\mathbbm{k}} + \,\overline\eth{\mathbbm{k}})
             - \mathbbm{b}\,\overline\eth\,\overline{\mathbbm{k}} - \,\overline{\mathbbm{b}}\,\eth{\mathbbm{k}}\,\right\}  \nonumber \\
             {}&+ \mathbb{F}_{\mathbb{K}}
             =  0 \label{eq:eth_constr_mom2} \\
     \partial_\rho  {\mathbbm{k}}    {}&  - \tfrac12\,\widetilde{\mathbb{N}} \,(\,\overline{\eth}\, \mathbbm{k}) -\tfrac12\, \,\overline{\widetilde{\mathbb{N}}}
         \, (\eth\,\mathbbm{k}) + \,\widehat{\mathbb{N}}\,(\mathbb{K})^{-1}\left\{\,\boldsymbol{\kappa}\,(\eth\,\mathbb{K}) 
         - \,\,\mathbbm{d}^{-1}\,[\,(\mathbbm{a}\,\mathbbm{k}-\mathbbm{b}\,\overline{\mathbbm{k}})\,(\eth{\,\overline{\mathbbm{k}}}) \right.{}
          \label{eq:eth_constr_mom1} \nonumber \\ 
            {}& \left. + (\mathbbm{a}\,\overline{\mathbbm{k}} -\,\overline{\mathbbm{b}}\,\mathbbm{k} )\,(\eth{\mathbbm{k}}) \,]  \right\} 
         + \mathbbm{f}_{\mathbbm{k}} =  0 \,,  \\
           {}& \boldsymbol\kappa= (2\,\mathbb{K})^{-1}\left[\,
               \mathbbm{d}^{-1}(  2\,\mathbbm{a}\,{\mathbbm{k}}\,\overline{\mathbbm{k}} - \mathbbm{b}\,\overline{\mathbbm{k}}^2
              - \,\overline{\mathbbm{b}}\,\mathbbm{k}^2)  - \tfrac12\,\mathbb{K}^2 - \boldsymbol\kappa_0 \,\right]\,,
              \label{eq:eth_constr_ham_n} 
\end{align}
where, in virtue of (\ref{constr_ham_n0}), $\boldsymbol\kappa_0$ can be evaluated by means of (\ref{R3}), (\ref{R2}),
(\ref{tracefreebfcurv_sq}), (\ref{hatextcurv_trace}), (\ref{hatextcurv_sq}) and (\ref{Lie_hatK}).

\medskip

In (\ref{eq:eth_constr_mom2})--(\ref{eq:eth_constr_mom1}), the lower order forcing terms $\mathbb{F}_{\mathbb{K}}$
and $\mathbbm{f}_{\mathbbm{k}}$ are spin-weight $0$ and $1$ fields, respectively,
on the level surfaces of the $\mycal{S}_\rho$ foliation. They are both smooth undifferentiated functions of the constrained variables
$\boldsymbol{\kappa},\mathbb{K}, \mathbbm{k}$; and they are also smooth functions of the freely specifiable variables
$\mathbbm{a},\mathbbm{b},  \,\widehat{\mathbb{N}}, \mathbb{N}, \interior{\mathbb{K}}$ and their various $\eth$, $\overline{\eth}$
and $\rho$ derivatives. The explicit form of the forcing terms is
\begin{align}
            \mathbb{F}_{\mathbb{K}} = {} &	\tfrac14\,\widehat{\mathbb{N}}\,\mathbbm{d}^{-1}\left\{    
	         2 \, \mathbbm{a} \,\mathbbm{B} \,\overline{\mathbbm{k} }  
		 -  \mathbbm{b}\, (\, \overline{ \mathbbm{C}} \,\mathbbm{k}
		 +\overline{ \mathbbm{A}} \, \overline{\mathbbm{k}} \,)
		+ ``\,CC\,"  \right \}_{\tiny(\ref{divk})} \\		
		&  -\mathbbm{d}^{-1} \left[ (\,
		\mathbbm{a} \,\overline  {\mathbbm{k}} -\overline {\mathbbm{b}}\, \mathbbm{k}\,)\,\eth\,\widehat{\mathbb{N}} 
		+ ``\,CC\," \right]_{\tiny(\ref{ndotk})} 
	 +\widehat{\mathbb{N}} \left[\,\interior{\rm\bf K}{}_{ij}   {\widehat K}{}^{ij}-(\,\boldsymbol{\kappa} -\tfrac12 \,\mathbb{K}\,)\,\widehat{\mathbb{K}}\,\right]_{\tiny(\ref{hatextcurv_sq})} \nonumber \\ 
            \mathbbm{f}_{\mathbbm{k}} = {}& -\tfrac12 \left[\,  \mathbbm{k} \,\eth \,\overline{\widetilde{\mathbb{N}}}
        +\overline {\mathbbm{k}}\, \eth \,{\widetilde{\mathbb{N}}} \right]_{\tiny(\ref{liek})} \\ {}& 
            +\tfrac12 \,\widehat{\mathbb{N}}\,(\mathbbm{d}\,\mathbb{K})^{-1} \left [\,
        (\mathbbm{a}\,\mathbbm{k}-\mathbbm{b}\,\overline{\mathbbm{k}})\,(\overline{\mathbb{B}}\,\mathbbm{k}+\mathbb{B}\,\overline{\mathbbm{k}}) 
         + (\mathbbm{a}\,\overline{\mathbbm{k}} -\,\overline{\mathbbm{b}}\,\mathbbm{k} )\,(\mathbb{C}\,\overline{\mathbbm{k}}
         +\mathbbm{A}\,\mathbbm{k})\, \right ]_{\tiny(\ref{2kDk})}  \nonumber \\ {}& 
            - [\,\boldsymbol\kappa-\tfrac12\, \mathbb{K}\,]\,  \eth\,\widehat{\mathbb{N}} 	    
            + \widehat{\mathbb{N}} \left[\,\tfrac12\,\mathbb{K}^{-1} \,\eth \boldsymbol\kappa_0 
        + \widehat{\mathbb{K}} \,\mathbbm{k} -q^i \dot{\widehat n}{}^l\,\interior{\rm\bf K}_{li}  
        + q^i\widehat D^l \interior{\rm\bf K}{}_{li}  \right]_{\tiny(\ref{qndotK})-(\ref{qDK})}   \nonumber
\end{align}
where the terms with parenthetical sub-indices are obtained by referring to the designated equations. 

\section{Final remarks}

A chief motivation for these rather heavy calculations is to yield evolution equations which can be integrated numerically in the radial $\rho$-direction as a coupled system of ordinary differential equations (ODEs), e.g.~by applying the method of lines to a finite difference or a pseudo-spectral representation of the $\eth$ and $\overline\eth$ operators, as described in \cite{jeff_edth,jeff_edth_2, maciej}.  
Note, however, that other  numerical methods can be applied to integrate  (\ref{eq:eth_constr_mom2})--(\ref{eq:eth_constr_mom1}).
For instance, the following spectral method may be preferable in various circumstances. 

\medskip

This method is based upon the spectral expansion of the spin-weighted fields
$\mathbbm{a},\mathbbm{b}, \,\widehat{\mathbb{N}}, \mathbb{N},  \interior{\mathbb{K}}; \boldsymbol{\kappa},$ $\mathbb{K}, \mathbbm{k}$ 
on the $\mycal{S}_\rho$ level surfaces, which can be expressed in the general pattern 
\begin{equation}\label{decomp0}
\mathbbm{x}= \sum_{l,m}\, x^{\,l,m}(\rho)\cdot {}_{s}\mathbb{Y}_{\,l,m}\,,
\end{equation}
where $\mathbbm{x}$ has spin-weight $s$ and ${}_{s}\mathbb{Y}_{\,l,m}$ stands for the corresponding spin-weighted spherical harmonics. 
In particular, $\mathbb{K}$ and $\mathbbm{k}$ have the decompositions 
\begin{equation}\label{decomp}
\mathbb{K}= \sum_{l,m}\, K^{\,l,m}(\rho)\cdot {}_{0}\mathbb{Y}_{\,l,m}\, \quad {\rm and} \quad
\mathbbm{k} = \sum_{l,m}\, k^{\,l,m}(\rho)\cdot {}_{1}\mathbb{Y}_{\,l,m}\, .
\end{equation}
Accordingly, after using (\ref{eq:eth_constr_ham_n}) to substitute for $\boldsymbol{\kappa}$,
(\ref{eq:eth_constr_mom2})--(\ref{eq:eth_constr_mom1}) become a system of coupled ODEs
for the expansion coefficients $K^{\,l,m}(\rho)$ and $k^{\,l,m}(\rho)$. This system can be then
be implemented numerically and integrated, e.g.~by means of a suitable adaptation of the numerical package GridRipper \cite{gridripper3,LARI} or by that of the method described in \cite{beyer}.

\medskip

Given the resulting solution for $\mathbb{K}$ and $\mathbbm{k}$, their substitution back into
(\ref{eq:eth_constr_ham_n}) yields $\boldsymbol{\kappa}$ on $\Sigma$ and thereby the full set of constrained  variables.
It then follows from Theorem 4.3 of \cite{racz_constraints} that the complete initial data $h_{ij}$ and $K_{ij}$ determined from the
constrained  variables and the freely specifiable part of the initial data is guaranteed to satisfy the constraints
(\ref{new_expl_eh})-(\ref{new_expl_em}) in the ``domain of dependence'' of $\mycal{S}_{{\rho}_{0}}$ in $\Sigma$.

\medskip

Recall, in this procedure, that the initial data for $\mathbb{K}$ and $\mathbbm{k}$ must also be freely specified
on the level surface $\rho=\rho_\circ$. The particular choice 
\begin{equation}\label{decomp_schw}
{K}^{\,l,m}= \left\{ \begin{array} {r l}  -\frac{8\,\sqrt{\pi}\,M}{{\rho_\circ}^2\,\sqrt{1+2\frac{M}{{\rho_\circ}}}}\,, &
{\rm if}\ l=m=0;\\ 0 , & {\rm otherwise}  \end{array}
\right.
\,,\quad 
{k}^{\,l,m}=0 \ {\rm for}\ \forall\ l, m  
\end{equation}
yields the Schwarzschild initial data, provided that all the freely specifiable functions take their associated
Schwarzschild values, as specified in \cite{i_jeff}.

\section*{Acknowledgments}

IR were supported in part by the Die Aktion \"Osterreich-Ungarn, Wissenschafts- und Erziehungskooperation grant 90\"ou1
and by the NKFIH grant K-115434. JW was supported by NSF grant PHY-1505965 to the University of Pittsburgh.
Both of the authors are grateful for the kind hospitality of
the Albert Einstein Institute in Golm, Germany,  where this work was initiated.  

\appendix
\section*{Appendix}\label{Appendix_A}
\renewcommand{\theequation}{A.\arabic{equation}}
\renewcommand{\thelemma}{A.\arabic{lemma}}

\setcounter{equation}{0}

This appendix verifies the basic relations between the operators $\eth$, $\overline{\eth}$ and the covariant derivative operator $\mathbb{D}_a$ on  $\mathbb{S}^2$. 

\medskip

To do so denote by $\mathbb{D}_a$ and $\partial_a$ the torsion free covariant derivative operators with respect to the metrics $q_{ab}$ and $\delta_{ab}$ in (\ref{conf_flat2}), respectively. They are related to the Christoffel symbols by
\begin{equation}
{\Gamma^e}{}_{ab}= \tfrac12\,q^{ef}\left\{ \partial_{a}q_{fb}+ \partial_{b}q_{af}-\partial_{f}q_{ab}\right\} 
     =  \Omega^{-1}\left\{ 2\,\delta^{e}{}_{(a}\partial_{b)}\Omega  -\delta_{ab}\,\delta^{ef}\partial_{f}\Omega\right\}\,. 
\end{equation} 
Using (\ref{conf_factor}) it is straightforward to verify that
\begin{equation}\label{conex}
q^aq^b \, {\Gamma^e}{}_{ab}= P \left\{ 2\,P^2\,\delta^{e}{}_{\,\overline z}\left(\partial_{\,\overline z}P^{-1}\right) \right\}= -2\,q^e\left(\partial_{\,\overline z}P\right) 
\end{equation}
\begin{equation}\label{conex2}
\,\overline q^aq^b\,  {\Gamma^e}{}_{ab}= P \left\{ P\left[q^{e}\left(\partial_{z}P^{-1}\right)+ \,\overline q^{e}\left(\partial_{\,\overline z}P^{-1}\right)\right]
       - \left(\,\overline q^aq^b\, q_{ab}\right) \,q^{ef}\left(\partial_{f} P^{-1}\right)\right\}=0 \, , 
\end{equation}
where in the last step of (\ref{conex2}) we used the relation
\begin{align}\label{conex3}
\hskip-0.0cm
\left(\,\overline q^aq^b\, q_{ab}\right) q^{ef}\left(\partial_{f} P^{-1}\right)= {}& 2\left[\tfrac12 \left(q^e\,\overline q^f+q^f\,\overline q^e\right)\right]\left(\partial_{f} P^{-1}\right) 
=  P\left[q^e\left(\partial_{z}P^{-1}\right) + \,\overline q^e\left(\partial_{\,\overline z}P^{-1}\right) \right]  
\end{align}
along with 
\begin{equation}\label{partial}
q^a\partial_a q^b = P\,\delta^{b}{}_{\,\overline z}\left(\partial_{\,\overline z}P \right)=q^b\left(\partial_{\,\overline z}P\right)
\end{equation}
and
\begin{equation}\label{partial2}
\,\overline q^a\partial_a q^b = P\,\delta^{b}{}_{\,\overline z}\left(\partial_{z}P \right)=q^b\left(\partial_{z}P\right)\,, 
\end{equation}
whereas in verifying  (\ref{conex3}) we used the normalization condition (\ref{unit_metr-norm}), along with the definition (\ref{dyad}).

\begin{lemma} 
As a consequence of (\ref{eth-def1})-(\ref{eth-def3})
\begin{align}
\eth\,\mathbb{L} = {}& q^b q^{a_1}\dots q^{a_s}\, \mathbb{D}_b\mathbf{L}_{({a_1}\dots{a_s})}\,,\\
\,\overline\eth\,\mathbb{L} = {}& {\,\overline q}^b q^{a_1}\dots q^{a_s}\, \mathbb{D}_b\mathbf{L}_{({a_1}\dots{a_s})}\,.
\end{align}
\end{lemma}
{\rm\bf Proof:} As
\begin{equation}
\mathbb{D}_b\mathbf{L}_{({a_1}\dots{a_s})}=\partial_b\mathbf{L}_{({a_1}\dots{a_s})}-\sum_{i=1}^s {\Gamma^e}{}_{ba_{i}}\,\mathbf{L}_{({a_1}\dots e_{_{_{_{\hskip 
	-.24cm{i\atop\smile}}}}}\hskip -.1cm \dots {a_s})}\, ,
\end{equation}
we have
\begin{align}
q^b q^{a_1}\dots q^{a_s}\, \mathbb{D}_b\mathbf{L}_{({a_1}\dots{a_s})}= {}& \bigl\{\, q^b\partial_b\, \mathbb{L}
         - \sum_{i=1}^s (q^b\partial_b  q^{a_i})\, q^{a_1} \dots \square _{_{_{_{\hskip 
	-.31cm{i\atop\smile}}}}}\hskip -.1cm \dots q^{a_s} \,\mathbf{L}_{({a_1}\dots {a_i} \dots{a_s})} \,\bigr\} \nonumber \\
{}& \phantom{q^a\partial_a} -\sum_{i=1}^s \left( {\Gamma^e}{}_{ba_{i}}\,q^bq^{a_i}\right)\,\mathbf{L}_{({a_1}\dots e_{_{_{_{\hskip 
	-.24cm{i\atop\smile}}}}}\hskip -.1cm \dots {a_s})} \,q^{a_1} \dots \square _{_{_{_{\hskip 
	-.31cm{i\atop\smile}}}}}\hskip -.1cm \dots q^{a_s} \,,
\end{align}
where $\square _{_{_{_{\hskip 
	-.31cm{i\atop\smile}}}}}$ indicates that the omission of the $i^{th}$ dyad element.

This, in virtue of (\ref{dyad}), (\ref{conex}) and (\ref{partial}), implies that 
\begin{align}
q^b q^{a_1}\dots q^{a_s}\, \mathbb{D}_b\mathbf{L}_{({a_1}\dots{a_s})}= {}& \left\{ P\left(\partial_{\,\overline z}\mathbb{L} \right)- s \left(\partial_{\,\overline z} P \right)\mathbb{L}\right\} + 2 s \left(\partial_{\,\overline z} P \right)\mathbb{L} \nonumber \\
= {}& P\left(\partial_{\,\overline z}\mathbb{L} \right)+ s \left(\partial_{\,\overline z} P \right)\mathbb{L} = P^{1-s}\,\partial_{\,\overline z} \left( P^s\,\mathbb{L} \right)= \eth\,\mathbb{L}\,.
\end{align}

By replacing $q^b$ by $\,\overline q^b$, and $\eth$ by $\,\overline\eth$ in the above argument,
the application of (\ref{dyad}), (\ref{conex2}) and (\ref{partial2}) yields the analogous relation 
\begin{equation}
\,\overline{q}^b q^{a_1}\dots q^{a_s}\, \mathbb{D}_b\mathbf{L}_{({a_1}\dots{a_s})}
    =P\left(\partial_{z}\mathbb{L} \right)- s \left(\partial_{z} P \right)\mathbb{L}
       = P^{1+s}\,\partial_{z} \left( P^{-s}\,\mathbb{L} \right)=\,\overline\eth\,\mathbb{L} \, ,
\end{equation}
as intended to be shown. \hfill $\square$


\end{document}